\documentclass[twocolumn,secnumarabic,amssymb, nobibnotes, aps, prd]{revtex4}

\newcommand{\be}{\begin{eqnarray}}
\newcommand{\ee}{\end{eqnarray}}

\def\nubar{{\bar{\nu}}}

\newcommand{\ms}{\Delta m^2_{21}}

\def\gs{\mathrel{
   \rlap{\raise 0.511ex \hbox{$>$}}{\lower 0.511ex \hbox{$\sim$}}}}
\def\ls{\mathrel{
   \rlap{\raise 0.511ex \hbox{$<$}}{\lower 0.511ex \hbox{$\sim$}}}}
\newcommand{\bea}{\begin{equation} \begin{array}{c}}
\newcommand{\bead}{\begin{equation} \begin{array}{cccc}}
\newcommand{\eea}{ \end{array} \end{equation}}
\usepackage{amssymb}
\usepackage{amsmath}
\usepackage{graphicx}
\usepackage{hyperref}
\usepackage{subfigure}
\usepackage{gensymb}
\usepackage{color}
\def\slc#1{\setbox0=\hbox{$#1$}           
    \dimen0=\wd0                                 
    \setbox1=\hbox{/} \dimen1=\wd1               
    \ifdim\dimen0>\dimen1                        
       \rlap{\hbox to \dimen0{\hfil/\hfil}}      
       #1                                        
    \else                                        
       \rlap{\hbox to \dimen1{\hfil$#1$\hfil}}   
       /                                         
    \fi}

\newcommand{\beq}{\begin{equation}}
\newcommand{\eeq}{\end{equation}}
\newcommand{\beqa}{\begin{eqnarray}}
\newcommand{\eeqa}{\end{eqnarray}}

\begin{document}

\title{Invisible neutrino decay in the light of NOvA and T2K data }
\author{Sandhya Choubey}
\email{sandhya@hri.res.in}
\affiliation{Harish-Chandra Research Institute, HBNI, Chhatnag Road, Jhunsi, Allahabad 211 019, India}
\affiliation{Department of Physics, School of
Engineering Sciences, KTH Royal Institute of Technology, AlbaNova
University Center, 106 91 Stockholm, Sweden}

\author{Debajyoti Dutta}
\email{debajyotidutta@hri.res.in}
\affiliation{Harish-Chandra Research Institute, HBNI, Chhatnag Road, Jhunsi, Allahabad 211 019, India}

\author{Dipyaman Pramanik}
\email{dipyamanpramanik@hri.res.in}
\affiliation{Harish-Chandra Research Institute, HBNI, Chhatnag Road, Jhunsi, Allahabad 211 019, India}

\begin{abstract}
We probe for evidence of invisible neutrino decay in the latest NOvA and T2K data.  It is seen that both NOvA and T2K data sets are better fitted when one allows for invisible neutrino decay. We consider a scenario where only the third neutrino mass eigenstate $\nu_3$ is unstable and decays into invisible components. The best-fit value for the $\nu_3$ lifetime is obtained as $\tau_{3}/m_{3} = 3.16\times 10^{-12}$ s/eV from the analysis of the NOvA neutrino data and  $\tau_{3}/m_{3} = 1.0\times 10^{-11}$  s/eV from the analysis of the T2K neutrino and anti-neutrino data. The combined analysis of NOvA and T2K gives $\tau_{3}/m_{3} = 5.01\times 10^{-12}$ s/eV as the best-fit lifetime. However, the statistical significance for this preference is weak with the no-decay hypothesis still allowed at close to 1.5$\sigma$ C.L. from the combined data sets, while the two experiment individually are consistent with no-decay even at the 1$\sigma$ C.L.  At 3$\sigma$ C.L., the NOvA and T2K data give a lower limit on the neutrino lifetime of $\tau_{3}/m_{3}$ is  $\tau_{3}/m_{3} \geq 7.22 \times 10^{-13}$ s/eV and $\tau_{3}/m_{3} \geq 1.41 \times 10^{-12}$ s/eV, respectively, while NOvA and T2K combined constrain $\tau_{3}/m_{3}  \geq  1.50 \times 10^{-12}$  s/eV. We also show that in presence of decay the best-fit value in the $\sin^{2}\theta_{23}$ vs $\Delta m^{2}_{32}$ plane changes significantly and the allowed regions increase significantly towards higher $\sin^{2}\theta_{23}$.

\end{abstract}

\maketitle

\section{Introduction}
The neutrino oscillation physics has entered its precision age.  Among the six oscillation parameters associated with the standard neutrino oscillation physics, $\ms$ and $\theta_{12}$ have been measured precisely from the solar neutrino experiments \cite{Ahmad:2002jz} and KamLAND \cite{Gando:2010aa}. The atmospheric neutrino experiments \cite{Fukuda:1998mi} first showed evidence of $\nu_{\mu} - \nu_{\tau}$ flavor transformation and gave a measurement of $\vert\Delta m_{32}^2\vert$ and $\theta_{23}$. The mixing angle $\theta_{13}$ is also now precisely determined from reactor experiments DayaBay \cite{An:2012eh}, Double CHOOZ \cite{Abe:2011fz} and RENO \cite{Ahn:2012nd}.  Currently the unknown quantities in the neutrino oscillation physics are the neutrino mass hierarchy/ordering {\it i.e.}, whether the lightest neutrino state is $\nu_{1}$ or $\nu_{3}$, the precise measurement of $\theta_{23}$ and its octant and the CP-violating phase $\delta_{CP}$. NOvA  \cite{Adamson:2016tbq, Adamson:2017gxd} and T2K \cite{Abe:2017uxa} are the presently running long baseline experiments. These are complimentary to the previous experiments and are expected to shed light on the unknown neutrino parameters.

The NOvA long-baseline experiment is in USA. The $\nu_{\mu}$  flux is generated by the NuMI beam at Fermilab. The experiment employs two identical Totally Active Scintillator Detector (TASD) which are different only in terms of their mass. The Near detector (ND)  is a 100 m deep 290 ton detector while the Far Detector (FD) is on the surface with mass 14 kton. The neutrino flux is measured first at the ND located at 1 km away from the target. The neutrino beam is next detected at the FD situated 810 km away near Ash River, Minnesota at an off-axis angle of 14.6 mrad. The 14.6 mrad off-axis location of the FD gives a narrow energy spectrum at the FD with a peak near 2 GeV, which is tuned close to the first oscillation maximum at this baseline. 

T2K (Tokai to Kamioka) is a long-baseline experiment in Japan. Here the $\nu_{\mu}$  beam is produced at the J-PARC accelerator complex in Japan by impinging a 30 GeV proton beam onto a carbon target. This experiment also employs a two detector set-up that are off-axis compared to the beam direction. The neutrino produced from the beam are detected first at the ND, ND280, at 280 m from the target. The FD is the Super-Kamiokande detected with fiducial mass 22.5 kton and situated 295 km away from the source at $2.5^\circ$ off-axis from the main beam axis, giving a narrow beam peaked at around 600 MeV, which for T2K's baseline corresponds to the first oscillation maximum.

NOvA and T2K have both presented their initial results. The T2K experiment announced their first result with $1.43\times10^{20}$ protons on target (POT) on electron appearance in 2011 \cite{Abe:2011sj}. With six observed electron candidate events and 1.5 expected backgrounds, T2K gave the first direct evidence of non-zero $\theta_{13}$ at $2.5\sigma$ C.L. The first results announcement on muon disappearance came a year later in 2012  \citep{Abe:2012gx} with the same POT of $1.43\times10^{20}$ and gave a best-fit of $\Delta m^{2}_{32}=2.65\times10^{-3}$ eV$^{2}$ and $\sin^{2}2\theta_{23} = 0.98$. The first  anti-neutrino result from T2K was published in \citep{Abe:2015ibe}, where they used $\bar\nu_\mu$ beam with $4.01\times10^{20}$ POT and obtained best fit of $\sin^{2}\theta_{23} = 0.45$ and $\Delta m^{2}_{32} = 2.51\times10^{-3}$ eV$^{2}$, very consistent with the measurements obtained using the $\nu_\mu$ beam. The T2K collaboration has published their results periodically since the first announcements and their data-sets and best-fit oscillation parameters have remained consistent, with $\theta_{23}$ close to maximal.  The latest result form T2K \citep{Abe:2017vif} used $7.482\times10^{20}$ POT for neutrino mode and $7.471\times10^{20}$ POT for anti-neutrino data. This currently gives the best fit $\Delta m^{2}_{32} = 2.52\pm0.08(2.51\pm0.08)\times 10^{-3}$ eV$^{2}$ and $\sin^{2}\theta_{23} = 0.55^{+0.05}_{-0.09}(0.55^{+0.05}_{0.08})$ for normal(inverted) ordering, using both electron appearance as well as muon disappearance data.
The first result of muon-neutrino disappearance from NOvA came in 2016 \citep{Adamson:2016xxw}, where they used $2.74\times10^{20}$ POT and got the best fit $\Delta m^{2}_{32} = (2.52^{+0.20}_{-0.18})\times10^{-3}$ eV$^{2}$ and $\sin^{2}\theta_{23}=0.43$ and 0.60.  The was immediately followed with first result on electron appearance data \citep{Adamson:2016tbq} with the same exposure. Next disappearance data came in 2017 \citep{Adamson:2017qqn} which used $6.05\times10^{20}$ POT and gave $\Delta m^{2}_{32} = (2.67 \pm 0.11)\times10^{-3}$ eV$^{2}$, while for $\sin^{2}\theta_{23} $ they obtained two statistically degenerate values $0.404^{+0.030}_{-0.022}$ and $0.624^{+0.022}_{-0.030}$ and claimed that the NOvA data disfavours maximal mixing at 2.6$\sigma$. Results from the combined analysis of NOvA's appearance and disappearance data was presented in \citep{Adamson:2017gxd}. So while T2K prefers maximal mixing for $\theta_{23}$, the early analysis from the NOvA showed $2.6\sigma$ preference for non-maximal mixing. This tension between the two experiments led several authors to propose new physics ideas to explain the tension between the two datasets. However, the NOvA collaboration has recently done an improved re-analysis of their disappearance dataset \cite{Radovic:2018xyz}. The newer analysis mainly addresses better the energy resolution of the hadron sample leading to an improved neutrino energy resolution. They have divided the muon events into four quantiles of different resolutions from  $\sim6\%$ to $\sim12\%$, based on their hadronic energy fraction. This approach changed the measurement of $\theta_{23}$ at NOvA with maximal $\theta_{23}$ mixing being preferred by NOvA as well. Hence, the tension between NOvA and T2K has been resolved for now and the datasets seem to be consistent with standard three-generation flavor oscillations.

Although the data appears to be consistent with the standard expected three-generation paradigm, there can still be new physics effects present in these experiments. One such new physics scenario is neutrino decay.  The active neutrino state could decay into another lighter active neutrino state and boson(s), or it could decay into a sterile fermion and boson(s). The former scenario is called visible neutrino decay \cite{Kim:1990km, Acker:1992eh,Lindner:2001fx}.(since the final state fermion is active and hence ``visible" to the detector) while the latter is known as invisible neutrino decay (since the final state fermion is sterile and hence ``invisible" to the detector). The bosonic state(s) are assumed to be invisible in both class of models. In this paper we will consider neutrino decay into all invisible states only. There are two possibilities for such models: (i) The neutrino could decay into a Dirac fermion \cite{Acker:1991ej,Acker:1993sz} $\nu_{j} \rightarrow \nubar_{iR} + \chi$, where $\nubar_{iR}$ is a right-handed singlet  and $\chi$ is an iso-singlet scalar carrying a lepton number. (ii) The neutrinos could decay into a Majorana fermion and a Majoron $\nu_{j} \rightarrow \nu_{s} + J$ \cite{Gelmini:1980re,Chikashige:1980ui}. To evade the constraints of the Z decay to invisible particles from LEP data, the Majoron should be dominantly singlet \cite{Pakvasa:1999ta}.  First idea of decaying neutrinos was proposed very early in order to explain the solar neutrino problem \cite{Bahcall:1972my}. Later neutrino oscillations along with decay solution to the solar neutrino problem was  studied in \cite{Acker:1993sz, Berezhiani:1991vk, Berezhiani:1992xg, Choubey:2000an, Bandyopadhyay:2001ct, Joshipura:2002fb, Bandyopadhyay:2002qg, Picoreti:2015ika}. These studies obtained bound on $\tau_{2}$ by considering $\nu_{2}$ to be unstable. The bound on $\tau_{2}$ from the solar data is $\tau_{2}/m_{2} > 8.7\times 10^{-7}$ s/eV at 99 \% C.L. \cite{Bandyopadhyay:2002qg}. (See \cite{Berryman:2014qha,Picoreti:2015ika} for a more recent study). The bound on $\tau_2$ from SN1987A are much more stringent  \cite{Frieman:1987as}. Atmospheric and long-baseline (LBL) neutrinos give the bound on the $\nu_{3}$ lifetime. To solve the atmospheric neutrino problem a pure neutrino decay was proposed in \cite{LoSecco:1998cd}, however this gave a very poor fit to the data. Authors of \cite{Barger:1998xk,Lipari:1999vh} considered neutrino decay with mixing and claimed that it could somewhat reproduce the SK results, however, the zenith angle dependent SK data gave a poor fit \cite{Fogli:1999qt}. In \cite{Barger:1998xk,Fogli:1999qt} the $\Delta m^{2}$ dependent terms were averaged out. However, if the unstable neutrino state is allowed to decay into a sterile state with which it does not mix then the constraints on $\Delta m^2$ can be relaxed. Two scenarios have been studied in this context. In \cite{Barger:1999bg} $\Delta m^{2} << 10^{-4}$ eV$^{2}$ was considered and it was claimed to fit the data better, however the analysis by the SK collaboration showed that this gives poorer fit than the only oscillation case \cite{Ashie:2004mr}. 
The second scenario was considered in \cite{Choubey:1999ir}, where $\Delta m^{2}$ was kept unconstrained. For SK data, it was shown that a small non-zero decay parameter and $\Delta m^2  \sim 0.003$ eV$^{2}$ gave a better fit to the data. The global analysis of atmospheric and MINOS data was performed in \cite{GonzalezGarcia:2008ru}. Although only oscillation gave the best-fit, but the decay plus oscillation scenario also gave a good fit.  The bound obtained from the analysis is $\tau_{3}/m_{3} \geq 2.9 \times 10^{-10}$ s/eV at the 90 \% C.L. Prospects of constraining the $\nu_3$ lifetime with atmospheric neutrino events at INO was studied in \cite{Choubey:2017eyg}. 

The case for visible neutrino decay in long-baseline experiments was considered for T2K in \cite{Gago:2017zzy} and DUNE in \cite{Coloma:2017zpg}.  However, the visible neutrino decay scenario is very tightly constrained from data from other experiments. Therefore, we consider in this paper only the case for invisible neutrino decay.  The analysis \cite{Gomes:2014yua} considered MINOS and T2K data with two generation of neutrinos and gave the bound, $\tau_{3}/m_{3} \geq 2.8 \times 10^{-12}$ s/eV at 90 \% C.L. The expected results at DUNE in the invisible neutrino decay scenario was worked out in \cite{Choubey:2017dyu}.

In this paper we present the current constraints on $\tau_{3}/m_{3}$ from the recent data from NOvA and T2K, using the full three generation oscillation framework with matter effects. We also study the effect of decay on the measurement of $\theta_{23}$ and $\Delta m^{2}_{32}$. As was shown in  \cite{Choubey:2017dyu}, there can be significant impact on the measured value of $\theta_{23}$ if decay is present. We will study how presence of decay changes the best-fit values of $\theta_{23}$ and $\Delta m^{2}_{32}$ as well as the C.L. contours allowed by the T2K and NOvA data. 

The paper is organised as follows. In the next section we give the theoretical description of neutrino propagation when the $\nu_3$ state decays into a sterile state with which it does not mix. Also given in section \ref{sec:theory} are the details of our simulation framework needed for the real data analysis of T2K and NOvA. We next give our result in section \ref{sec:result} and finally we conclude in section \ref{sec:concl}.

\section{Invisible neutrino decay and simulation framework}\label{sec:theory}

We assume that the $\nu_{3}$ state decays into a sterile neutrino and a singlet scalar ($\nu_{3} \rightarrow \bar{\nu_{4}} + J$).We also assume that the neutrino mass eigenstate decays, therefore the mass matrix as well as the decay matrix can be simultaneously diagonalised. In this case the mixing between the flavor and the mass eigenstates can be written as,
\beq
\begin{pmatrix}\label{eq:mix}
\nu_{\alpha}\\
\nu_{s}
\end{pmatrix}
=
\begin{pmatrix}
U & 0\\
0 & 1
\end{pmatrix}
\begin{pmatrix}
\nu_{i}\\
\nu_{4}
\end{pmatrix},
\eeq
where the greek indices represent the standard flavor states {\it i.e.}, $e$, $\mu$, $\tau$ and the latin indices represent the mass eigenstates. $U$ is the standard PMNS matrix. Here we assume NH {\it i.e.}, $m_{3} > m_{2} > m_{1}$. In presence of decay the evolution equation in matter becomes:
\begin{widetext}
\beq\label{eq:evol}
i\frac{d}{dx}\begin{pmatrix}
\nu_e\\
\nu_\mu\\
\nu_\tau
\end{pmatrix}
 = \left[U\left[\frac{1}{2E}
\begin{pmatrix}
0&0&0\\
0&\Delta m^2_{21}&0\\
0&0&\Delta m^2_{31}
\end{pmatrix}
-i\frac{\alpha_3}{2E}
\begin{pmatrix}
0&0&0\\
0&0&0\\
0&0&1
\end{pmatrix}\right]U^\dagger
+ 
\begin{pmatrix}
      A&0&0\\0&0&0\\0&0&0
     \end{pmatrix}
\right]
\begin{pmatrix}
\nu_e\\
\nu_\mu\\
\nu_\tau
\end{pmatrix}
\,,
\eeq
\end{widetext}
where $A = 2\sqrt{2}G_{F}n_{e}E$ represents the matter potential due to neutrino electron scattering in matter, $G_{F}$ is the Fermi coupling, $E $ is the neutrino energy and $n_{e}$ is the electron density. We solve Eq.~(\ref{eq:evol}) numerically using Runge-Kutta with constant matter density. 

The simulation is done using a modified version of GLoBES, with modifications needed for real data analysis.

For the analysis of NOvA we have taken a 14 kt detector at a baseline of 812 km with constant matter density.  We have taken 8.5\% energy resolution for electron events and 6\% resolution for muon events.  The signal efficiency is chosen to be 99 \% for electron events and 91 \% for muon events. We normalize the number of events to match the best fit event spectra 
given in \citep{Adamson:2017gxd} for electrons ($6.04\times 10^{20}$ POT) and in \citep{Radovic:2018xyz} for muons ($8.85\times10^{20}$ POT). 


For the analysis of T2K, we have taken a 22.5 kt detector at a baseline of 295 km with constant matter density. The energy resolution is taken 8.5 \%.  The signal efficiency is chosen to be 51.5 \% for electron events and 90 \% for muon events.
We normalise the event spectra to match the event spectra given in  \citep{Abe:2017vif} which corresponds to $7.482\times10^{20}$ POT in neutrino mode and $7.471\times10^{20}$ POT for anti-neutrino mode.


\section{Results}\label{sec:result}

\begin{figure}
\includegraphics[width=0.45\textwidth]{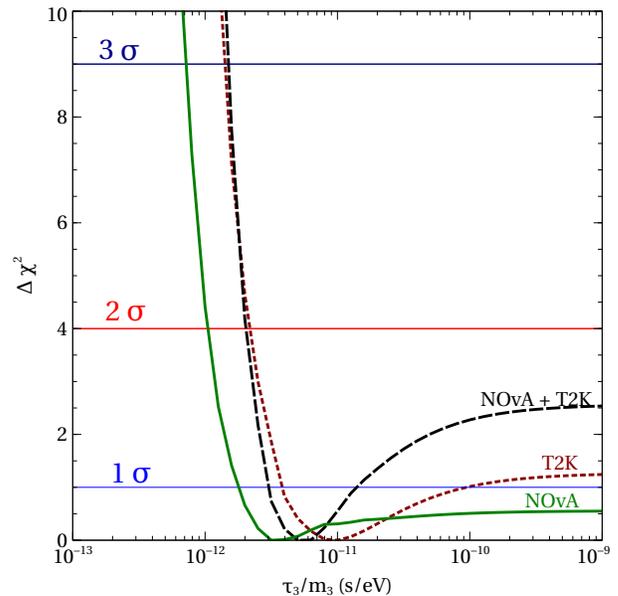}
\caption{\label{tau}
$\Delta \chi^2=\chi^2-\chi^2_{\rm min}$ vs $\tau_3/m_3$ obtained from the analysis of T2K data (red dashed line), NOvA data (solid green line) and T2K+NOvA data (black long dashed line).}
\end{figure}

\begin{figure}
\includegraphics[width=0.45\textwidth]{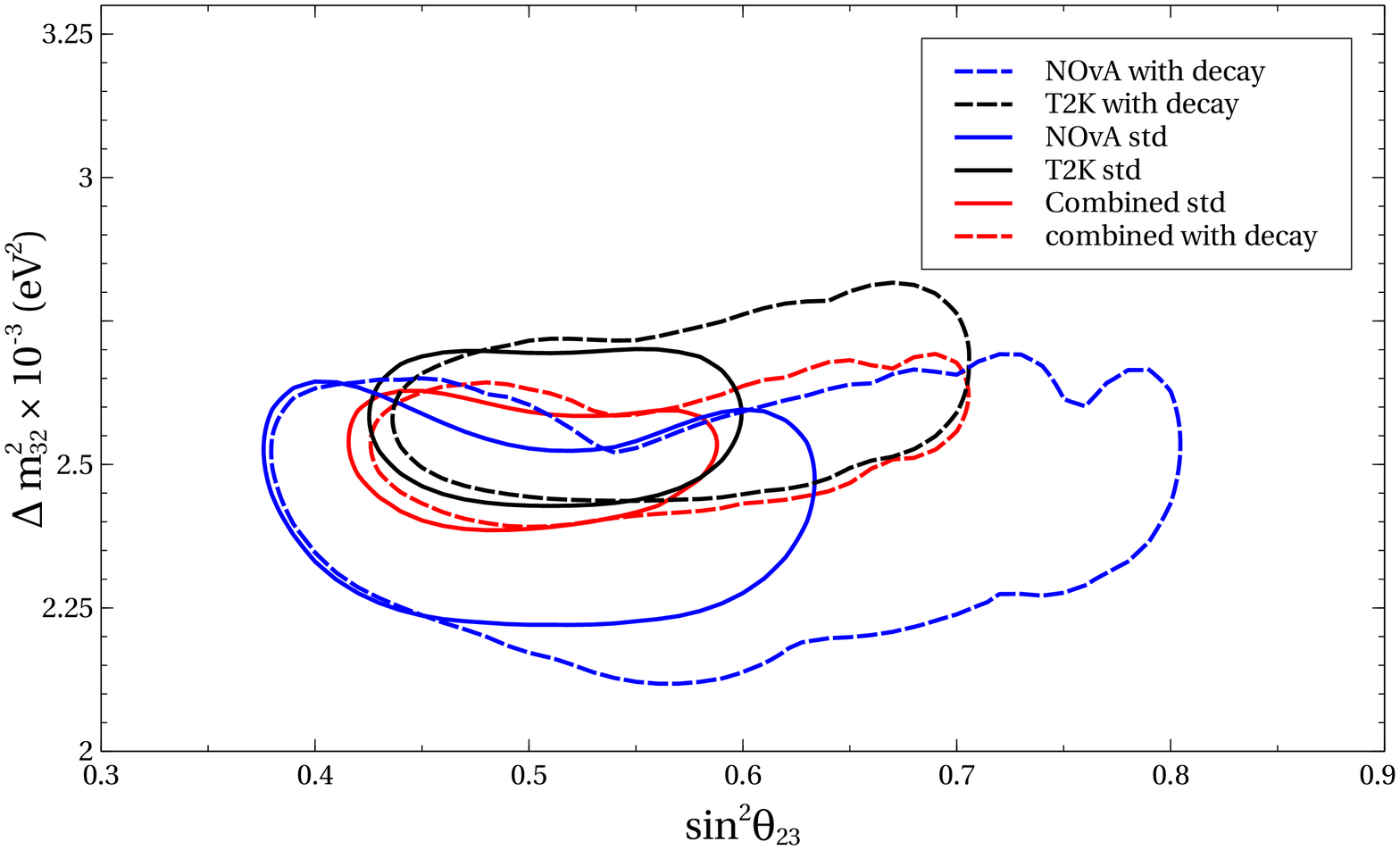}
\caption{\label{dm32th23}
The 95 \% C.L. allowed areas obtained in the $\sin^2\theta_{23}-\Delta m_{32}^2$ plane, from analysis of T2K data (red lines), NOvA data (green lines) and T2K+NOvA data (black lines). The solid lines are for standard three-generation oscillations while the dashed lines are for oscillation with decay. 
}
\end{figure}

In Fig.~\ref{tau} we show the constraint on $\tau_{3}/m_{3}$ from the current data of NOvA and T2K, where the latest appearance as well as disappearance data sets of both experiments have been taken into consideration in the analysis. The green solid curve is obtained using NOvA data alone, the dark red dotted curve is obtained using T2K data alone, while the black dashed curve is obtained from a combined analysis of NOvA and T2K data. The parameters $\theta_{23}$ and $\Delta m_{32}^2$ are marginalised in the fit over their 3$\sigma$ allowed ranges and $\delta_{CP}$ is marginalised over its full range. It can be seen from the Fig.~\ref{tau}, that for both experiments the no decay scenario is slightly disfavoured and the best-fit value from the fit comes for non-zero decay. NOvA disfavors the no decay scenario at 0.7$\sigma$ and the best-fit value is $\tau_{3}/m_{3} = 3.16\times10^{-12}$ s/eV. T2K disfavors the no decay case at slightly more than 1$\sigma$ and the best-fit value is $1\times10^{-11}$ s/eV. For the combined analysis case, the no decay case is disfavoured at 1.5 $\sigma$ and the best-fit obtained from the combination of the two data is $\tau_{3}/m_{3} =5.01 \times 10^{-12}$ s/eV. The minimum $\chi^{2}$ ($\chi^2_{\rm min}$) for NOvA, T2K and the combined case are 10.38, 69.34 and 87.19, respectively, which are slightly less than the standard oscillation fit, for which the $\chi^2_{\rm min}$ are 10.93, 70.39 and 88.65, respectively. Therefore, the  invisible decay scenario we consider in this work, fits the data slightly better than the standard oscillation case. The data sets also set a lower bound on the lifetime. The 3$\sigma$ lower bound on $\tau_{3}/m_{3}$ from NOvA data is seen to be $\tau_{3}/m_{3}\geq7.22\times 10^{-13}$ s/eV, while from T2K it is $\tau_{3}/m_{3} \geq 1.41 \times 10^{-12}$ s/eV. The 3$\sigma$ combined constraint from both experiments taken together is $\tau_{3}/m_{3} \geq 1.50 \times 10^{-12}$ s/eV. It can be clearly seen from Fig.~\ref{tau}, that the sum of the two $\Delta \chi^{2}$ is less than the $\Delta \chi^{2}$ for the combined analysis. This points at a synergy between the two experiments. This synergy results in an improved fit for the decay scenario compared to standard oscillation when we perform the combined analysis of the two experiments.

Fig.~\ref{dm32th23} shows the allowed region in the $\sin^{2}\theta_{23}$ vs $\Delta m^{2}_{32}$ plane at 95 \% C.L. The solid curves are for only oscillation case without decay whereas the dashed curves are for the case where $\nu_3$ are allowed to decay. The fit is marginalised over $\delta_{CP}$ in both cases and over $\tau_{3}/m_{3}$ as well for the case of decay plus oscillation. The blue curves are for NOvA, the black curves are for T2K and the red curves are for the combined analysis. For the standard case the best-fit points ($\sin^2\theta_{23}$, $\Delta m_{32}^2$) are (0.45, $2.41\times 10^{-3}$ eV$^2$), (0.52, $2.56\times 10^{-3}$ eV$^2$) and (0.46, $2.51\times 10^{-3}$ eV$^2$) for NOvA, T2K and the combined cases, respectively. On the other hand, for the case with decay and oscillation the corresponding best-fit points are: (0.48,$2.39\times 10^{-3}$ eV$^2$), (0.62, $2.62\times 10^{-3}$ eV$^2$) and (0.48, $2.52\times 10^{-3}$ eV$^2$) for NOvA, T2K and the combined cases, respectively. The interesting point to notice here is that for all cases, the allowed region of the parameter space increases significantly when decay is considered along with oscillation. Also note that with inclusion of decay, the best-fit shifts towards higher values of  $\sin^{2}\theta_{23}$. This behaviour is very similar to what was seen in \citep{Gomes:2014yua} in the context of MINOS and T2K and in \citep{Choubey:2017dyu} in the context of DUNE. 
The shift of $\theta_{23}$ to higher values in presence of decay can be understood in terms of the survival probability given in \citep{Gomes:2014yua} in the two-generation approximation neglecting matter effect, 
\begin{widetext}
\beq\label{eq:pmm_dec}
P^{2G}_{\mu\mu} = \left[\cos^{2}\theta_{23} + \sin^{2}\theta_{23}\exp(-m_{3}L/\tau_{3}E)\right]^{2}-\sin^{2}2\theta_{23}\exp(-m_{3}L/\tau_{3}E)\sin^{2}\Big(\frac{\Delta m^{2}_{31}L}{4E}\Big)
\,.
\eeq
\end{widetext}
In Eq.~\eqref{eq:pmm_dec}, there is an exponential suppression due to neutrino decay in both the oscillatory as well as the non-oscillatory term.  Therefore for a given $\theta_{23}$, the survival probability for the decay case will be less than the standard oscillation case. Hence, when decay is considered in the fit, the value of $\sin^2\theta_{23}$ increases in order to reproduce the same probability obtained for the standard case. 


\begin{figure}
\includegraphics[width=0.45\textwidth]{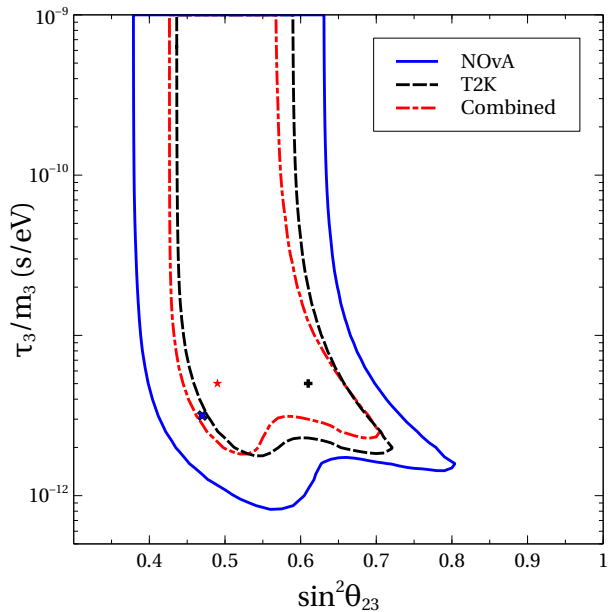}
\caption{\label{sintau}
The 95 \% C.L. allowed areas obtained in the $\tau_3/m_3 - \sin^2\theta_{23}$ plane, from analysis of T2K data (black dashed line), NOvA data (blue solid line) and T2K+NOvA data (red dashed-dotted line).}
\end{figure}

\begin{figure*}
\includegraphics[width=0.3\textwidth]{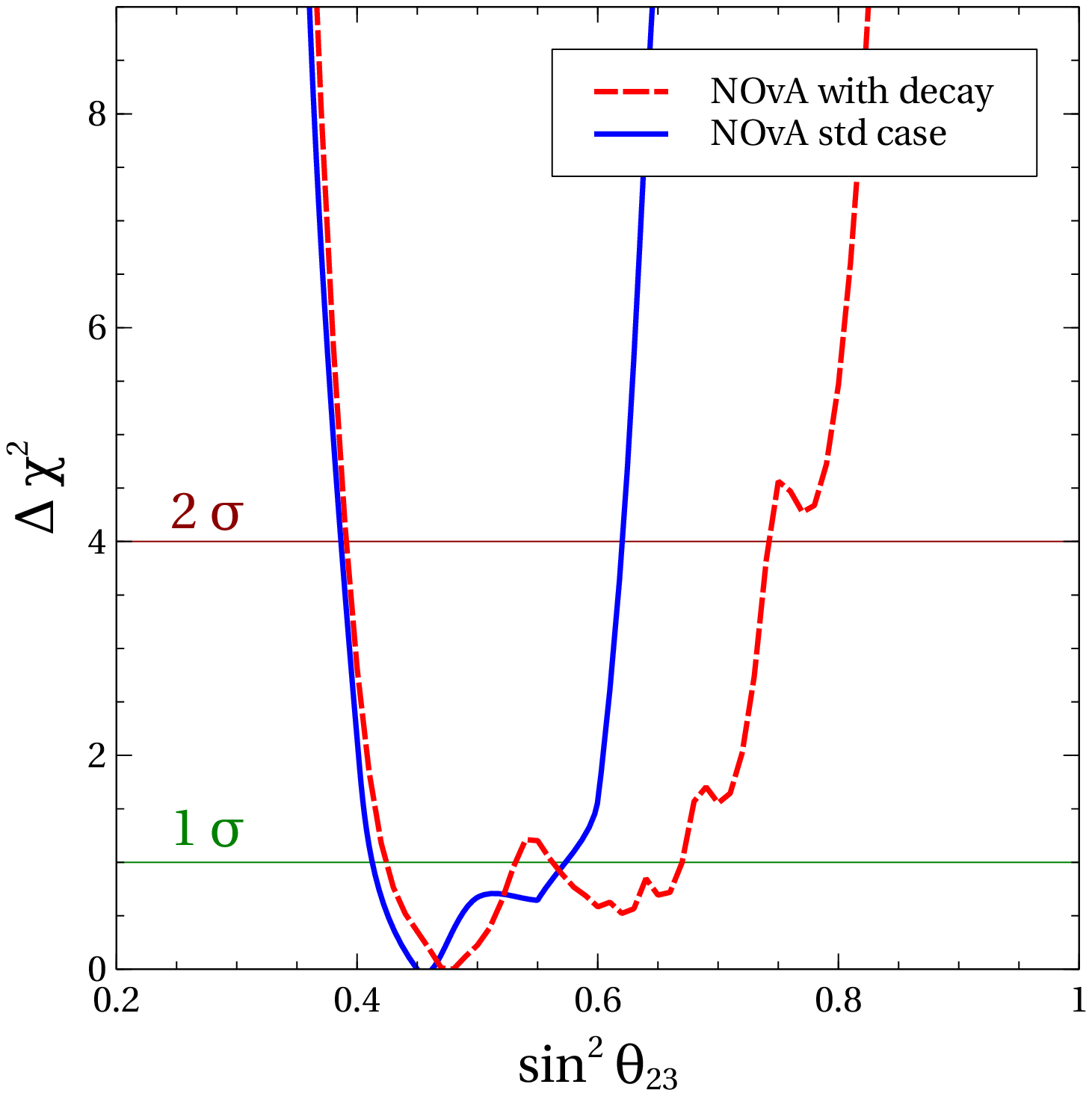}\quad
\includegraphics[width=0.3\textwidth]{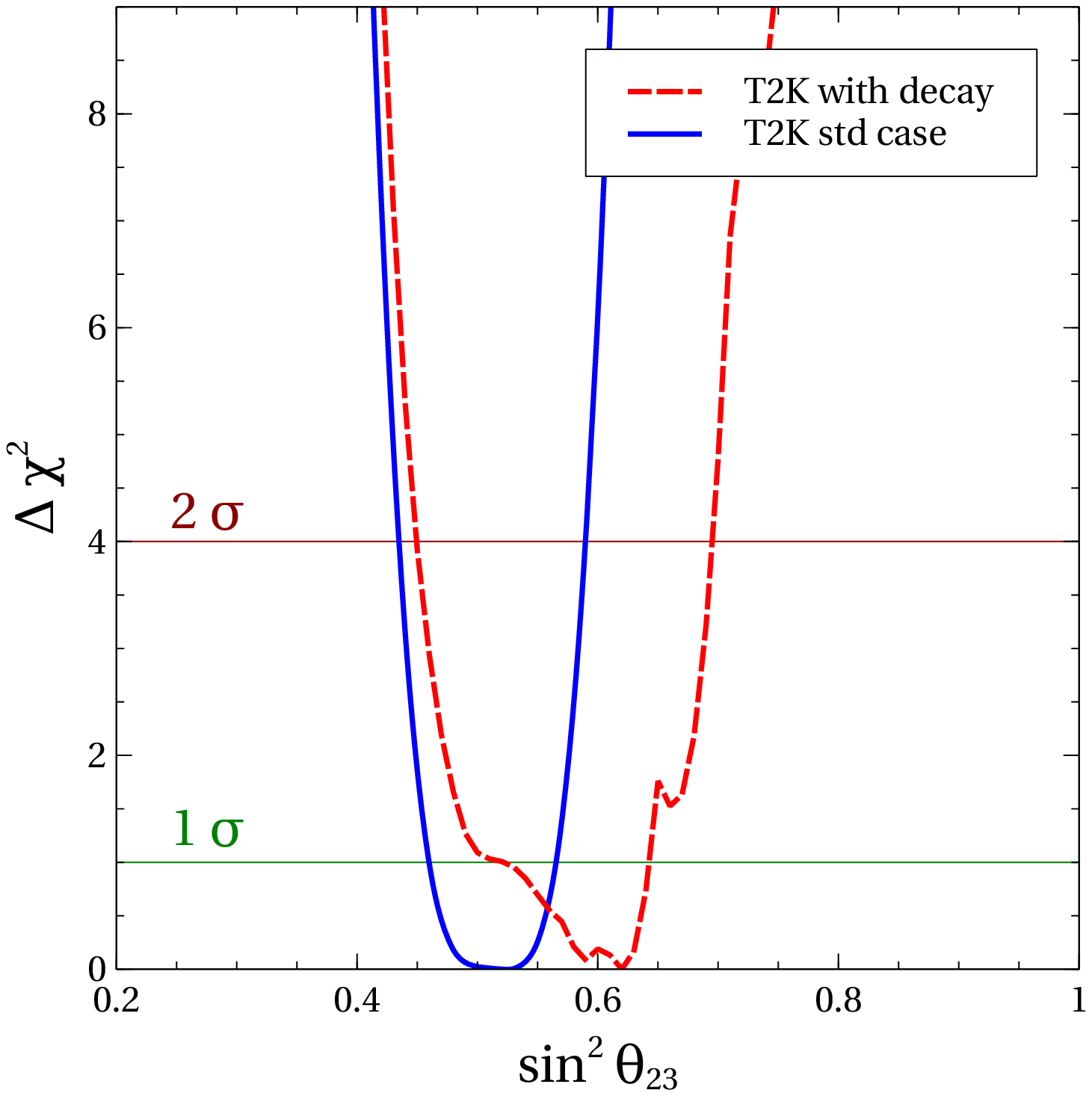}\quad
\includegraphics[width=0.3\textwidth]{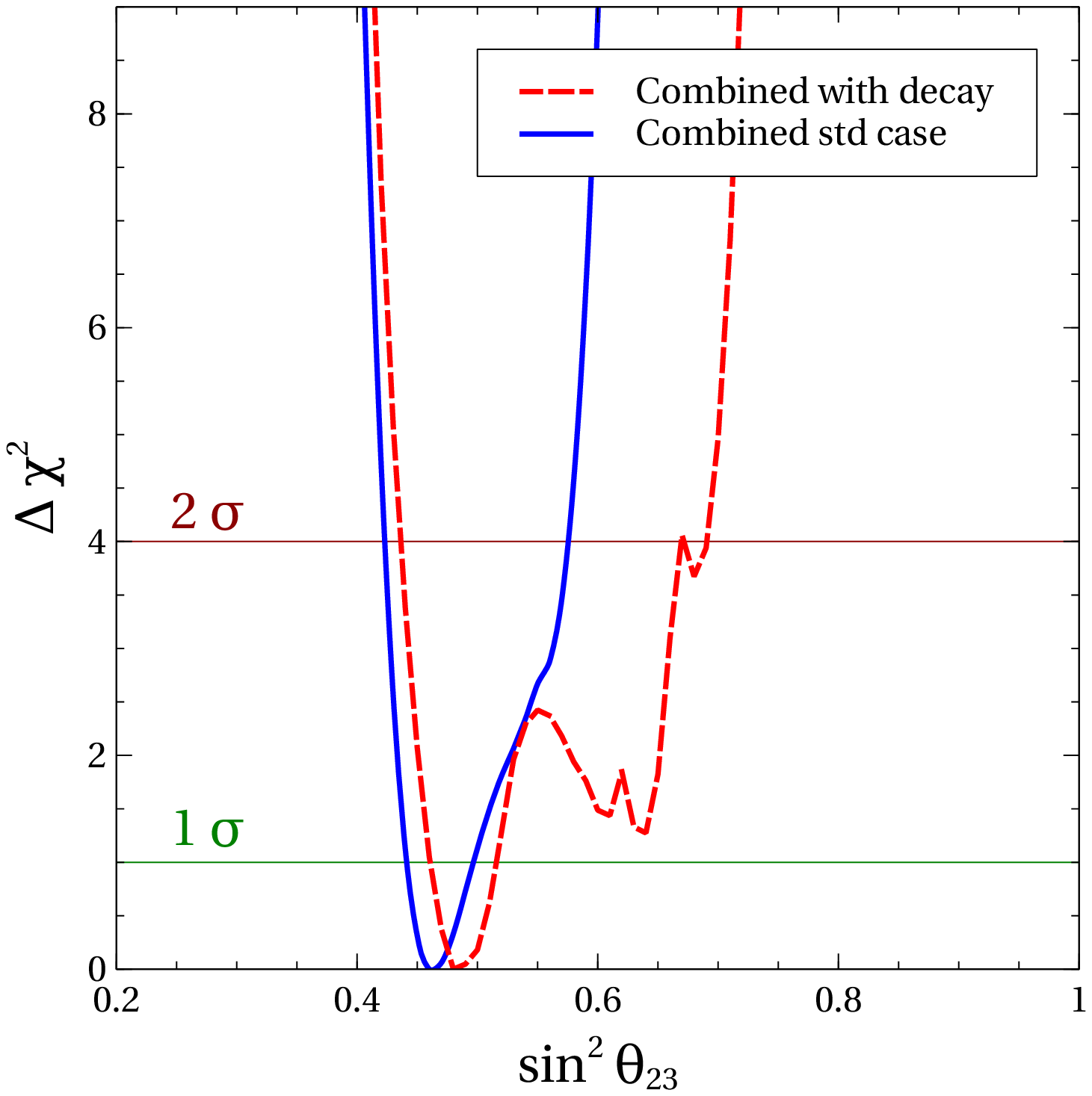}

\caption{\label{s23}
$\Delta \chi^2$ vs $\sin^{2}\theta_{23}$, where the $\chi^2$ is marginalised over all other parameters.}
\end{figure*}

Fig.~\ref{sintau} gives the allowed region in the $\sin^{2}\theta_{23}$ vs $\tau_{3}/m_{3}$ plane at 95 \% C.L. with $\Delta m_{32}^2$ marginalised over its current 3$\sigma$ range and $\delta_{CP}$ marginalised over full range. The blue solid, black dashed and red dashed-dotted curves are for NOvA, T2K and the combined analysis, respectively. The blue cross gives the best-fit for NOvA (0.47, $3.16 \times 10^{-12}$ s/eV), the black plus gives the best-fit for T2K (0.61, $5.011 \times 10^{-12}$ s/eV) and the red star gives the best-fit for the combined case (0.49, $5.011 \times 10^{-12}$ s/eV). Again as in Fig.~\ref{dm32th23}, the best-fit is seen to be for finite $\tau_{3}/m_{3}$.

Fig.~\ref{s23} gives the $\Delta \chi^{2}$ vs $\sin^{2}\theta_{23}$ for the standard case and the decay plus oscillation case. The fit is marginalised over $\Delta m_{32}^2$ and $\delta_{CP}$ for the standard case and  over $\Delta m_{32}^2$, $\tau_{3}/m_{3}$ and $\delta_{CP}$ for the decay plus oscillation case. The left panel is for NOvA, the middle panel is for T2K and the right panel is for the combined analysis. In all the panels the blue solid curves are for the standard oscillation case while the red dashed curves represent the decay with oscillation case. For T2K our standard oscillation best-fit $\sin^2\theta_{23}=0.52$ matches very well with the best-fit obtained by the T2K collaboration \cite{Abe:2017vif}. For NOvA on the other hand, our best-fit for standard oscillation comes at $\sin^2\theta_{23}=0.45$ while the NOvA collaboration gets their best-fit at $\sin^2\theta_{23}=0.558^{+0.041}_{-0.033}$ in the higher octant. 
The reason for this mild mis-match could be because our experimental simulation is based on GLoBES which cannot include all systematics in a rigorous manner. However, the figure shows that the $\chi^{2}$ differences between the minima of the two octant is less than 1, and hence the best-fit obtained by us for the higher octant is not so different from the correct value obtained from the collaboration. 

\begin{figure}
\includegraphics[width=0.45\textwidth]{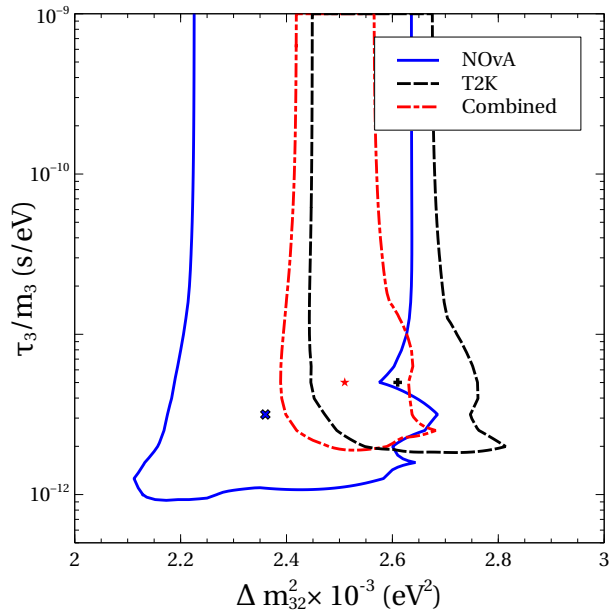}
\caption{\label{dm32tau}
The 95 \% C.L. allowed areas obtained in the $\tau_3/m_3 - \Delta m_{32}^2$ plane, from analysis of T2K data (black dashed line), NOvA data (blue solid line) and T2K+NOvA data (red dashed-dotted line).
}
\end{figure}

Fig.~\ref{dm32tau} gives the allowed region in the $\Delta m^{2}_{32}$ vs $\tau_{3}/m_{3}$ plane at 95 \% C.L. with $\theta_{23}$ marginalised over current 3$\sigma$ allowed region and $\delta_{CP}$ marginalised over full range. The blue curve is for NOvA, the black is for T2K and the red is for the combined case. The blue cross gives the best-fit for NOvA ($2.36 \times 10^{-3}$, $3.16 \times 10^{-12}$), the black plus for the T2K ($2.61 \times 10^{-3}$, $5.011 \times 10^{-12}$) and the red star is for the combined best-fit ($2.51 \times 10^{-3}$, $5.011 \times 10^{-12}$). The above best-fit values are in units of (eV$^2$, s/eV).

\begin{figure*}
\includegraphics[width=0.3\textwidth]{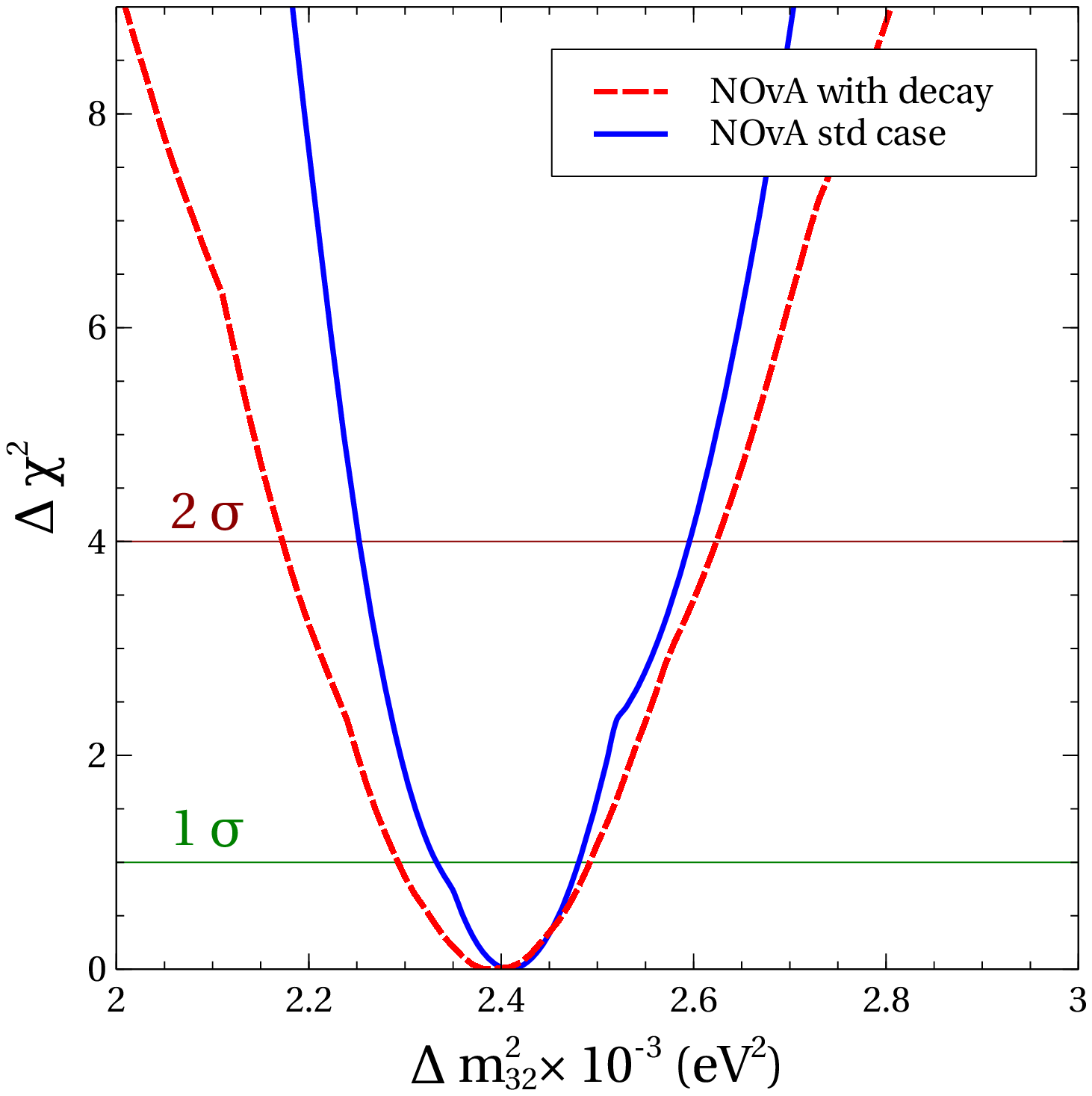}\quad
\includegraphics[width=0.3\textwidth]{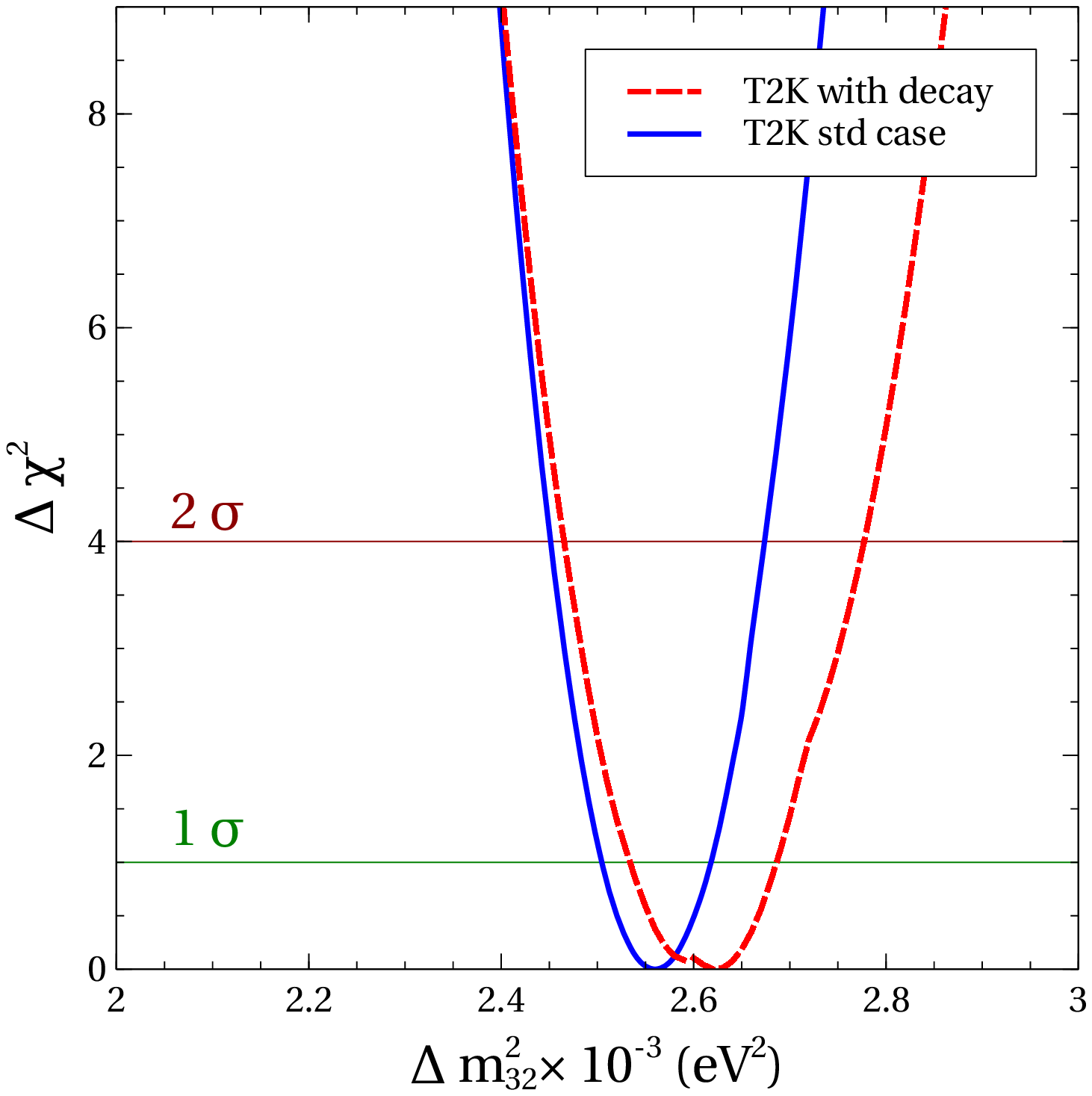}\quad
\includegraphics[width=0.3\textwidth]{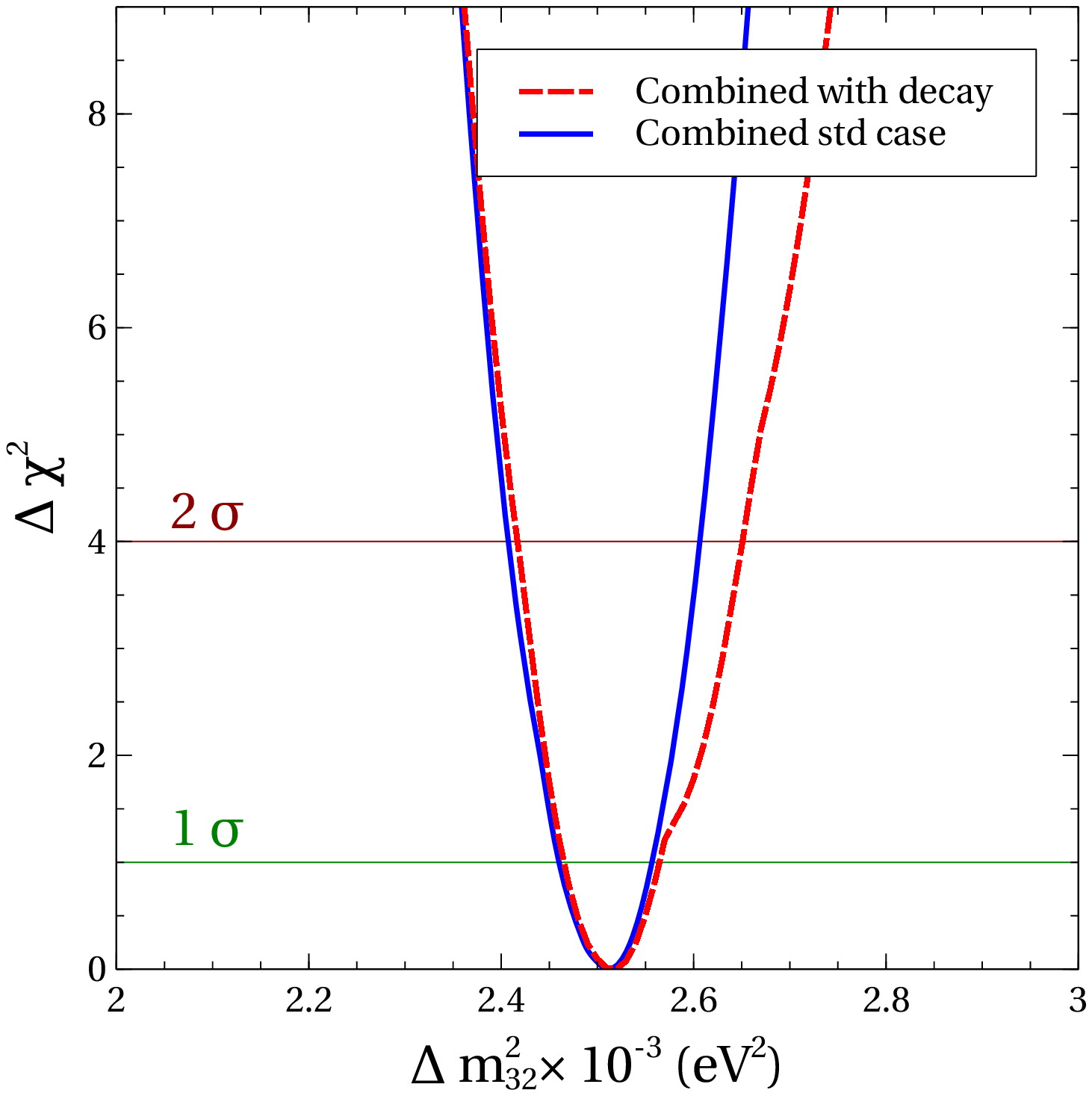}

\caption{\label{dm32}
$\Delta \chi^2$ vs $\Delta m^{2}_{32}$, where the $\chi^2$ is marginalised over all other parameters.}
\end{figure*}

Fig.~\ref{dm32} gives the $\Delta \chi^{2}$ vs $\Delta m^{2}_{32}$ with $\theta_{23}$ and $\delta_{CP}$ marginalised for the standard and $\theta_{23}$, $\delta_{CP}$ and $\tau_{3}/m_{3}$ marginalised for the case of decay plus oscillation. Just as in Fig.~\ref{s23}, the left panel is for NOvA, the middle is for T2K and the right panel is for the combined case.  The blue solid curves are for the standard case whereas the red dashed curves are for the decay plus oscillation case. It can be seen from the figure that for NOvA the allowed range of $\Delta m^{2}_{32}$ increases on both sides while for T2K the best-fit $\Delta m^{2}_{32}$ shifts towards higher values and as a result the allowed range shifts towards the right. 

\section{Summary \& conclusion}\label{sec:concl}
We analysed the recent NOvA and T2K data for invisible decay of the neutrino. We considered a framework where the $\nu_{3}$ is unstable and it decays into some lighter sterile state with which the active neutrinos do not mix. We used modified GLoBES for simulating the T2K and NOvA experiments taking into account the experimental exposure, systematic uncertainties and resolution functions for the current data sets. We considered the latest disappearance and appearance data given by the T2K collaboration which corresponds to $7.482\times10^{20}$ POT for neutrino and $7.471\times10^{20}$ POT for anti-neutrinos \cite{Abe:2017vif}. For NOvA we consider the disappearance data and simulation framework announced by the collaboration in \cite{Radovic:2018xyz} and appearance data presented in \cite{Adamson:2017gxd}. To obtained the oscillation probabilities we considered the full three generation framework with matter effects and decay and solved the evolution equation for the neutrinos numerically. We performed a two-pronged analysis. On one hand we looked at how compatible the current T2K and NOvA data are with neutrino decay. On the other hand we checked how the best-fit values and allowed ranges on $\Delta m_{32}^2$ and $\sin^2\theta_{23}$ change when decay is included in the fit.

We found that both T2K and NOvA data give better fit when neutrino decay in included. The best-fit lifetime corresponding to the T2K data is $\tau_{3}/m_{3} = 1.0 \times 10^{-11}$ s/eV. This can be compared with the best-fit lifetime of $\tau_{3}/m_{3} = 1.6 \times 10^{-12}$ s/eV obtained in \cite{Gomes:2014yua} using the older T2K data \cite{Abe:2014ugx} which corresponded to $6.57\times10^{20}$ POT for the neutrino mode and did not have the anti-neutrino data. The main difference between the analysis performed in \cite{Gomes:2014yua} and this work is the use of the anti-neutrino data, more exposure in the neutrino data and the inclusion of the electron appearance data. Our best-fit can also be compared to the best-fit lifetime of $\tau_{3}/m_{3} = 1.2 \times 10^{-12}$ s/eV obtained in \cite{Gomes:2014yua} from the combined fit of T2K  and MINOS  data. The best-fit $\nu_3$ lifetime from NOvA data corresponds to $\tau_{3}/m_{3} = 3.16 \times 10^{-12}$ s/eV, which is about an order of magnitude smaller than the T2K best-fit. The combined fit of T2K and NOvA returns a best-fit $\tau_{3}/m_{3} = 5.01 \times 10^{-12}$ s/eV. The datasets also put a lower bound on the $\nu_3$ lifetime. The $3\sigma$ bound put by T2K, NOvA and T2K+NOvA are $\tau_{3}/m_{3} \geq 1.41 \times 10^{-12}$ s/eV, $\tau_{3}/m_{3} \geq 7.22 \times 10^{-13}$ s/eV and $\tau_{3}/m_{3} \geq 1.50 \times 10^{-12}$ s/eV, respectively. 

We also studied the effect of decay on the measurement of the standard parameters $\theta_{23}$ and $\Delta m_{32}^2$. We found that if we include decay in our fit, the best-fit values for $\sin^{2}\theta_{23}$ and $\Delta m^{2}_{32}$ change significantly. The best-fit ($\sin^2\theta_{23},\Delta m_{32}^2$) obtained for the standard oscillation case from analysis of NOvA, T2K and both experiments combined are (0.45, $2.41\times 10^{-3}$ eV$^2$), (0.52, $2.56\times 10^{-3}$ eV$^2$) and (0.46, $2.51\times 10^{-3}$ eV$^2$) respectively. On including decay in the fit, the corresponding best-fit points become (0.48,$2.39\times 10^{-3}$ eV$^2$) for NOvA, (0.62, $2.62\times 10^{-3}$ eV$^2$) for T2K and (0.48, $2.52\times 10^{-3}$ eV$^2$) for the NOvA and T2K combined. The best-fit $\sin^2\theta_{23}$ is seen to be shifting to higher values. We also give the 95 \% C.L. contours in the two-parameter space and the $\Delta \chi^2$ vs $\sin^2\theta_{23}$ and $\Delta \chi^2$ vs $\Delta m_{32}^2$ plots from which 1, 2 and 3$\sigma$ ranges of these parameters can be read for both hypothesis, with and without decay. Decay is seen to shift the allowed range of $\sin^2\theta_{23}$ significantly to higher values, thereby extending the allowed ranges in the higher octant. The reason for this behavior was discussed. The allowed range of $\Delta m_{32}^2$ is also seen to change with inclusion of delay, albeit very mildly. 

In conclusion, both T2K and NOvA, and in particular NOvA, seem to favor neutrino decay. Even though this conclusion is not statistically significant yet, it will be interesting to see the results from the forthcoming next-generation long baseline experiments like DUNE and T2HK. Invisible neutrino decay also results in shifting $\theta_{23}$ to higher values and this would be again an interesting phenomenon to study at the next-generation experiments.

\section*{Acknowledgement}
We acknowledge the HRI cluster computing facility (http://www.hri.res.in/cluster/). This project has received funding from the European Union's Horizon 2020 research and innovation programme InvisiblesPlus RISE under the Marie Sklodowska-Curie grant agreement No 690575. This project has received funding from the European Union's Horizon 2020 research and innovation programme Elusives ITN under the Marie Sklodowska- Curie grant agreement No 674896.

\bibliography{ref}

\end{document}